\documentclass[10pt]{article}

\begin{document}

\title{Self-similar cosmologies in $5D$: spatially flat anisotropic models}
\author{J. Ponce de Leon\thanks{E-mail: jpdel@ltp.upr.clu.edu or  jpdel1@hotmail.com}\\ Laboratory of Theoretical Physics, Department of Physics\\ 
University of Puerto Rico, P.O. Box 23343, San Juan, \\ PR 00931, USA} 
\date{July   2008}

\maketitle

\begin{abstract}

In the context  of theories of Kaluza-Klein type, with a large extra dimension, we study self-similar cosmological models in $5D$ that are homogeneous, anisotropic and spatially flat. The ``ladder"  to go between the physics in $5D$ and $4D$ is provided by Campbell-Maagard's embedding theorems. We show that the $5$-dimensional field equations $R_{AB} = 0$ determine the form of the similarity variable. There are  three different possibilities: homothetic, conformal and ``wave-like" solutions in $5D$.
We derive the most general homothetic and conformal solutions to the $5D$ field equations. They require the extra dimension to be spacelike, and are given  in terms of one arbitrary function of the similarity variable  and three parameters. The Riemann tensor in $5D$ is not zero, except in the isotropic limit, which corresponds to the case where the parameters are equal to each other. 
The solutions can be used as $5D$ embeddings for a great variety of $4D$ homogeneous cosmological models, with and without matter, including the Kasner universe.  
Since the extra dimension is spacelike, the $5D$ solutions are invariant under the exchange of spatial coordinates. Therefore they also embed  a family of spatially {\it inhomogeneous} models in $4D$. We show that these models can be interpreted as vacuum solutions in braneworld theory. Our work (I) generalizes the $5D$ embeddings used for the FLRW models; (II) shows that 
anisotropic cosmologies are, in general,  curved in $5D$,    in contrast with FLRW models which can always be embedded in a $5D$ Riemann-flat (Minkowski) manifold; (III)  reveals that anisotropic  cosmologies  can be curved and devoid of matter, both   in $5D$ and $4D$, even when the  metric in $5D$ explicitly depends on the extra coordinate, which is quite different from the isotropic case.

\end{abstract}

\medskip



\newpage
\section{Introduction}

In standard general relativity, exact solutions have played a crucial role in the development of many areas of astrophysics and cosmology \cite{Bicak}, \cite{MacCallum}. Exact solutions  (i) provide a route to better and deeper understanding of gravity by giving  non-perturbative insight into the highly nonlinear gravitational phenomena,  (ii) can reveal unforeseen   features of the theory that might be  relevant to more general situations, and (iii) can be used for checking computer codes, which is important for the advent of numerical relativity \cite{Yasadjiev}.

In recent years there has been an increased interest in theories that envision our world as embedded in a universe with more than four {\it large} extra dimensions. Besides the longstanding theoretical motivations - as, e.g.,  resolving the differences between gravity and quantum field theory, and unifying all forces of nature -  the  interest in these theories has now a {\it practical} stimulus. Namely, that any solution of the N-dimensional Einstein equations with source can be locally embedded in a $(N + 1)$-dimensional Ricci-flat manifold, which is guaranteed  by  Campbell-Maagard's embbedings theorems \cite{T1}-\cite{Indefenceof}.  In particular,  this means that  {\it any} solution of the $4D$ Einstein equations with matter $G_{\mu\nu} = 8 \pi T_{\mu \nu}$ (throughout the paper we set $c = G = 1$) can always be locally embedded in a solution of the vacuum Einstein field equations $R_{AB} = 0$ in $5D$. 

Finding an embedding for a particular $4D$ spacetime,  with a specified physical energy-momentum tensor,  is an interesting and important question. However, it faces a fundamental problem. Namely,  that the effective $4D$ equations for gravity  contain a source term, that is the spacetime projection of the $5D$ Weyl tensor, which is unknown without specifying the properties of the metric in $5D$ \cite{Maartens}. Therefore, the best way of deriving  some practical benefit from  Campbell-Maagard's theorem is by solving the equations in $5D$, under some general geometrical properties, and then study the possible interpretations as well as the characteristics  of the matter sources in $4D$ that can generate these properties.  

From a single solution in $5D$ one may construct a great variety of scenarios in $4D$ ranging from static configurations to cosmological solutions \cite{ExtraSymmetry}, which reflects the fact that (a) there are many ways of embedding a $4D$ spacetime in a given higher-dimensional manifold, and (b) that the effective equations is $4D$ do not form a closed system of equations. Therefore, the study of exact solutions of $5D$ vacuum Einstein's equations is essential;  not only because of the arguments (i)-(iii) given above, and  for the formal mathematical aspects associated with the theory (e.g., the classification of $5D$ manifolds) but also, for the growing importance of the application of Kaluza-Klein to cosmological and astrophysical phenomena in $4D$. 
  The study of all possible scenarios is important in order to be able to predict observable effects in $4D$ caused by new physics from an extra dimension.

Nearly all the solutions in $5D$ are obtained under the assumption of spatial spherical symmetry and  the choice of one or more  additional functional relations  that simplify the field equations in such a way that they become fully integrable. 
The disadvantage of this approach is that different choices may lead to the same physical solution, in different disguise. 

In a recent work \cite{UniverseTopSep} we started a systematic investigation of the solutions of the $5D$ field equations $R_{A B} = 0$ for the case where the $5D$ metric possesses self-similar symmetry. We found the most general self-similar, homogeneous and isotropic, Ricci-flat cosmologies in $5D$ and showed that they can be interpreted, or used, as $5D$ Riemann-flat embeddings for spatially flat FRW cosmologies in $4D$. In this paper we continue that investigation, but we abandon the assumption of perfect spatial isotropy and consider cosmologies with spatial anisotropy. 

For the models to be physically realistic they have to closely resemble the standard FRW models. It is well known that, within the context of Bianchi classification, the generalization of the  ``flat",  ``open" and ``closed" FRW models are cosmologies of Bianchi type I, V and IX, respectively \cite{Jacob}. Cosmological observations today indicate that our universe is spatially flat \cite{Bernandis}-\cite{Spergel2}. Thus, Bianchi I models are the closest  anisotropic approximation to the present stage of evolution of our universe. 

Consequently, in this paper we adopt the $5$-dimensional line element\footnote{In this work we use the following conventions: $ t = x^0,\; x = x^1,\; y = x^2$ and $z = x^ 3$ are the usual spacetime coordinates; $\psi = x^4$ represents the coordinate along the ``extra" dimension; the signature of the $5D$ metric is $(+, -, -, -, \epsilon)$ where $\epsilon$ can be either $- 1$ or $+ 1$ depending on whether the extra dimension is spacelike or timelike. The range of tensor indices is $A, B...= 0-4$ and $\mu, \nu, ... = 0-3$.} 
\begin{equation}
\label{metric for Bianchi I model in 5D}
dS^2 =  e^{\nu(t, \psi)}dt^2 - e^{\lambda(t, \psi)}dx^2 - e^{\mu(t, \psi)}dy^2 - e^{\sigma(t, \psi)}dz^2  + \epsilon e^{\omega(t, \psi)}d\psi^2, 
\end{equation}
which on every hypersurface $\psi$ = constant, reduces to  homogeneous and anisotropic  Bianchi type-I cosmological models with flat spatial sections. A simplified version of this metric has been studied by Roque and Seiler \cite{Roque} under the assumption of separability. In this work we obtain the general solutions to   the field equations $R_{A B} = 0$ under the assumption  that the metric (\ref{metric for Bianchi I model in 5D}) possesses  self-similar symmetry. This assumption is motivated not only by the fact that FRW models are self-similar in $4D$ \cite{JPdeL2} and $5D$ \cite{UniverseTopSep}, but also by the ``similarity hypothesis" of Carr and Coley \cite{Coley}, which asserts  that under a variety of physical circumstances, both homogeneous and inhomogeneous cosmological models (in $4D$) naturally evolve to a self-similar form\footnote{Although there are some counterexamples to this hypothesis \cite{Wainwright}, \cite{Apostolopoulos 4}, there is a strong evidence that self-similar models play a significant role at asymptotic regimes  \cite{Apostolopoulos 2}-\cite{Apostolopoulos 3}. }.

We demonstrate that the field equations in $5D$ allow three types of similarity solutions: homothetic, conformal and ``wave-like". 
We show how to integrate the field equations for the first two types of similarity. We find that the solutions  (a)
depend on one arbitrary function and three arbitrary parameters; (b) require the extra dimension to be spacelike, and (c) generalize  the family of homogeneous and isotropic $5D$ models  discussed in \cite{UniverseTopSep}. In accordance with Campbell-Maagard's theorem they can be used, or interpreted, as $5D$ embeddings for four-dimensional subspaces. Depending on how we do the actual embedding, one can construct $4D$ anisotropic cosmological models   that are either homogeneous  or inhomogeneous, although homogeneous isotropic cosmologies are  recovered as particular cases.  

By means of dimensional reduction of the metric  in $5D$, we analyze, in some detail, possible applications of the family of homothetic solutions in $5D$.  
We find that all the  models constructed in $4D$,   exhibit  some type of self-similarity.  
Those that are inhomogeneous and anisotropic inherit the homothetic symmetry from the $5D$ embedding. The rest of them are either partially homothetic or homothetic along some $4D$ vector. Thus, all these models represent different self-similar scenarios in $4D$ and,  as such,  can be  relevant to the similarity hypothesis mentioned above. 

The paper is organized as follows. In section $2$, we deduce the shape of the similarity variable.  In section $3$, we solve the field equations with homothetic and conformal symmetry in $5D$. In section $4$, we present some general properties of the solutions in $5D$. In section $5$ we show different admissible interpretations in $4D$, but an exhaustive treatment of all possibilities is beyond the scope of this work. In section $6$ we give a summary.

\section{Similarity variable}
In  a self-similar model,  by a suitable transformation of coordinates all the dimensionless quantities can be put in a form where they are functions only of a single variable (say $\zeta$) \cite{Sedov}-\cite{JPdeL5}. Thus, in the case under consideration, in ``self-similar" coordinates $(\bar{t}, \bar{x}, \bar{y}, \bar{z}, \bar{\psi})$, the line element (\ref{metric for Bianchi I model in 5D}) can be written as 

\begin{equation}
\label{sel-similar metric for Bianchi I model in 5D}
dS^2 =  e^{\nu(\zeta)}d\bar{t}^2 - e^{\lambda(\zeta)}d\bar{x}^2 - e^{\mu(\zeta)}d\bar{y}^2 - e^{\sigma(\zeta)}d\bar{z}^2  + \epsilon e^{\omega(\zeta)}d\bar{\psi^2}, 
\end{equation}
where $\zeta$ is some function of $\bar{t}$ and $\bar{\psi}$, viz.,
\begin{equation}
\label{general self-similar variable}
\zeta = \zeta(\bar{t}, \bar{\psi}).
\end{equation}
The metric functions $\nu(\zeta)$, $\lambda(\zeta)$, $\mu(\zeta)$, $\sigma(\zeta)$, $\omega(\zeta)$, as well as $\zeta$ and the signature coefficient $\epsilon$, have to satisfy the field equations $R_{AB} = 0$.  In order to avoid misunderstanding, it is worth mentioning that in the literature the concept of self-similarity is frequently equated with homothetic symmetry. In this work we follow the traditional nomenclature used in \cite{Sedov}-\cite{Barenblat}. Among other things, we will see that the self-similar line element  
(\ref{sel-similar metric for Bianchi I model in 5D}) does not necessarily admit a homothetic Killing vector. In what follows we are going to suppress the bar over the self-similar coordinates.

The shape of the similarity variable $\zeta$ is determined by the field equations. Firstly,  we note that from $R_{04} = 0$ it follows that  
\begin{equation}
\label{equation for the shape of the similarity variable}
2\left(\lambda_{\zeta} + \mu_{\zeta} + \sigma_{\zeta}\right)\dot{\zeta}' + \left[\left(\lambda_{\zeta}^2 + \mu_{\zeta}^2 + \sigma_{\zeta}^2\right) - \left(\lambda_{\zeta} + \mu_{\zeta} + \sigma_{\zeta}\right)(\nu_{\zeta}+ \omega_{\zeta}) + 2 \left(\lambda_{\zeta\zeta} + \mu_{\zeta\zeta} + \sigma_{\zeta\zeta}\right)\right]{\zeta}' \dot{\zeta} = 0,
\end{equation}
where $f_{\zeta}$ denotes derivative of $f$ with respect to $\zeta$; dots and primes stand for  derivatives with respect to $t$ and $\psi$, respectively. 

\subsection{Homothetic and conformal self-similar variable}

Equation (\ref{equation for the shape of the similarity variable}) implies that, in the general case where $\dot{\zeta}' \neq 0$, the assumed 
self-similarity requires that the ratio $[\dot{\zeta}'/({\zeta}' \dot{\zeta})]$ be some function of  $\zeta$. Clearly, {\it any} separable function of $t$ and $\psi$ will do the job. Therefore, without loss of generality  we can set
\begin{equation}
\label{general zeta}
\zeta = \frac{T({t})}{F({\psi})},
\end{equation}
where $T$ and $F$ are some functions of their arguments. Consequently, the equations  $R_{00} = 0, R_{11} = 0, R_{22} = 0, R_{33} = 0$ and $R_{44} = 0$ they all have the structure

\begin{equation}
\label{structure of the equations}
\left(\frac{{\dot{T}}^2}{F'^2}\right)M(\zeta) + \left(\frac{\ddot{T} F}{F'^2}\right)N(\zeta) + \left(2 -  \frac{FF''}{F'^2}\right)P(\zeta)+ Q(\zeta)   = 0,
\end{equation}
where $M$, $N$, $P$ and $Q$ symbolize the corresponding functions of $\zeta$ in these equations. Therefore, in order to preserve the self-similar symmetry,  we have to require 
\begin{equation}
2 -  \frac{FF''}{F'^2} = l,
\end{equation}
where $l$ is a dimensionless separation constant.
Integrating this expression we obtain
\begin{equation}
F' = p F^{(2 -l)},
\end{equation}
where $p$ is a constant of integration. Thus, we find 
\begin{equation}
\left(\frac{{\dot{T}}^2}{F'^2}\right) = \left[\frac{{\dot{T}}^2}{p^2 T^{(4 - 2l)}}\right]\zeta ^{(4 - 2l)},\;\;\;\mbox{and}\;\;\;\left(\frac{\ddot{T} F}{F'^2}\right) = \left[\frac{\ddot{T}}{p^2 T^{(3 - 2l)}}\right]\zeta^{(3 - 2l)}.
\end{equation}
Consistency of (\ref{structure of the equations}) demands the  quantities inside the square brackets to be constants, which requires
\begin{equation}
T \sim \left\{\begin{array}{cc}
            t^{1/(l - 1)}    & \mbox{for $l \neq 1$}, \\
\\
               e^{q t}  & \mbox{for $l = 1$},
               \end{array}
      \right.
\end{equation}
where $q$ is some constant\footnote{We note that $p$ and $q$ must have the units of (length)$^{-1}$.}. 
Consequently, without loss of generality we can set
\begin{equation}
\label{general choice of zeta}
\zeta = \left\{\begin{array}{cc}
            \left(\frac{t}{\psi}\right)^{1/(l - 1)}    & \mbox{for $l \neq 1$}, \\
\\
               \left(\frac{e^{q t}}{e^{p \psi}}\right)  & \mbox{for $l = 1$}.
               \end{array}
      \right.
\end{equation}
We will see that the variables $\zeta$ for  $l \neq 1$ and $l = 1$ lead to homothetic and conformal solutions in $5D$, respectively.
\subsection{Wave-like self-similar variable}

Finally, in the particular case where $\dot{\zeta}' = 0$, it is not difficult to show that the similarity variable must have the form 
\begin{equation}
\label{xi in the particular case}
\zeta = \omega_{0}t + k_{0} \psi,
\end{equation} 
where $\omega_{0}$ and $k_{0}$ are some constants with the appropriate units. Thus, in this case the metric functions have a dependence of time and the extra coordinate like in  traveling waves or pulses propagating along the fifth dimension. Planes waves or wave-like solutions in $5D$, in the case of spatial spherical symmetry,  have been studied by Wesson, Liu and Seahra \cite{LiuWesson}, \cite{WessonLiuSeahra}; Horowitz, Low and Zee \cite{Horowitz}; as well as by the present author \cite{JPdeLWaveSolutions}.

\section{Field equations}

The field equations $R_{AB} = 0$ look  quite different for $\dot{\zeta}' \neq  0$ and $\dot{\zeta}' = 0$. In what follows we will consider the general case where  $\dot{\zeta}' \neq  0$, for which the metric functions depend on (\ref{general choice of zeta}). Wave-like solutions will be considered elsewhere.

Let us notice that for $l = 2$ the similarity variable becomes $\zeta_{(l = 2)} \equiv \xi =  (t/\psi)$. Thus, for an arbitrary $l \neq 1$ we can write $\zeta = \xi^{1/(l - 1)}$. What this means is that we can first study the integration of the equations for $l = 2$, and then obtain the solution for any value of $l$ by a simple transformation 
\begin{equation}
\label{transforming the solutions from l = 2 to any l}
\xi \rightarrow \;\zeta^{l - 1}, \;\;\;\xi \equiv \frac{t}{\psi}.
\end{equation}

\subsection{Integrating the field equations for $l = 2$}

The non-vanishing components of the Ricci tensor in $5D$ are 
\begin{enumerate}
  \item $R_{00} = 0$, 
  \begin{eqnarray}
\label{R00}
\nu_{\xi}\left[\left(\epsilon e^{\omega} - \xi^2 e^{\nu}\right)\left(\nu_{\xi} + \lambda_{\xi} + \mu_{\xi} + \sigma_{\xi} + \frac{2 \nu_{\xi\xi}}{\nu_{\xi}}\right) - 4\xi e^{\nu} + \omega_{\xi}\left(\epsilon e^{\omega} + \xi^2 e^{\nu}\right)\right] = \nonumber \\
\epsilon \left[\left(\nu_{\xi}^2 + \lambda_{\xi}^2 + \mu_{\xi}^2 + \sigma_{\xi}^2 + \omega_{\xi}^2\right) + 2\left(\nu_{\xi \xi} + \lambda_{\xi \xi} + \mu_{\xi \xi} + \sigma_{\xi \xi} + \omega_{\xi \xi}\right)\right]e^{\omega}. 
\end{eqnarray}

\item $R_{11} = 0$, 
\begin{equation}
\label{R11}
\lambda_{\xi}\left[\left(\epsilon e^{\omega} + \xi^2 e^{\nu}\right)\left(\lambda_{\xi} + \frac{2 \lambda_{\xi\xi}}{\lambda_{\xi}} + \mu_{\xi} + \sigma_{\xi}\right) + 4\xi e^{\nu} + \left(\epsilon e^{\omega} - \xi^2e^{\nu}\right)\left(\omega_{\xi} - \nu_{\xi}\right)\right] = 0.
\end{equation}

\item $R_{22} = 0$,
\begin{equation}
\label{R22}
\mu_{\xi}\left[\left(\epsilon e^{\omega} + \xi^2 e^{\nu}\right)\left(\mu_{\xi} + \frac{2 \mu_{\xi\xi}}{\mu_{\xi}} + \lambda_{\xi} + \sigma_{\xi}\right) + 4\xi e^{\nu} + \left(\epsilon e^{\omega} - \xi^2e^{\nu}\right)\left(\omega_{\xi} - \nu_{\xi}\right)\right] = 0.
\end{equation}

\item $R_{33} = 0$,
\begin{equation}
\label{R33}
\sigma_{\xi}\left[\left(\epsilon e^{\omega} + \xi^2 e^{\nu}\right)\left(\sigma_{\xi} + \frac{2 \sigma_{\xi\xi}}{\sigma_{\xi}} + \lambda_{\xi} + \mu_{\xi}\right) + 4\xi e^{\nu} + \left(\epsilon e^{\omega} - \xi^2e^{\nu}\right)\left(\omega_{\xi} - \nu_{\xi}\right)\right] = 0.
\end{equation} 

\item $R_{44} = 0$,
\begin{eqnarray}
\label{R44}
\omega_{\xi}\left[\left(\epsilon e^{\omega} - \xi^2 e^{\nu}\right)\left(\lambda_{\xi} + \mu_{\xi} + \sigma_{\xi} + \omega_{\xi} + \frac{2\omega_{\xi \xi}}{\omega_{\xi}}\right) - \nu_{\xi}\left(\epsilon e^{\omega} + \xi^2 e^{\nu}\right) - 4 \xi e^{\nu}\right] = \nonumber \\
- \xi\left[4\left(\nu_{\xi} + \lambda_{\xi} + \mu_{\xi} + \sigma_{\xi} + \omega_{\xi}\right) + \xi \left(\nu_{\xi}^2 + \lambda_{\xi}^2 + \mu_{\xi}^2 + \sigma_{\xi}^2 + \omega_{\xi}^2\right) + 2\xi \left(\nu_{\xi \xi} + \lambda_{\xi \xi} + \mu_{\xi \xi} + \sigma_{\xi \xi} + \omega_{\xi \xi}\right)\right]e^{\nu}.
\end{eqnarray}

\item Finally, $R_{04} = 0$,

\begin{equation}
\label{R04}
2\xi\left(\lambda_{\xi\xi} + \nu_{\xi \xi} + \sigma_{\xi \xi}\right) + \left(2 - \xi\nu_{\xi} - \xi \omega_{\xi}\right)\left(\lambda_{\xi} + \sigma_{\xi} + \mu_{\xi}\right) + \xi\left(\lambda_{\xi}^2 + \mu_{\xi}^2 + \sigma_{\xi}^2\right) = 0.
\end{equation}

\end{enumerate}

We note that equations (\ref{R11}), (\ref{R22}) and (\ref{R33}) require
\begin{equation}
\label{compatibility condition}
\frac{\lambda_{\xi\xi}}{\lambda_{\xi}} = \frac{\mu_{\xi \xi}}{\mu_{\xi}} = \frac{\sigma_{\xi \xi}}{\sigma_{\xi}}.
\end{equation}
Therefore, without loss of generality one can set
\begin{equation}
\label{introduction of f}
e^{\lambda} = A f^{2 \alpha}(\xi); \;\;\;e^{\mu} = B f^{2 \beta}(\xi), \;\;\;e^{\sigma} = C f^{2 \gamma}(\xi),
\end{equation}
where $A, B, C$ are constants;  $f$ is some function of the variable $\xi = (t/\psi)$;  and  $\alpha$, $\beta$ and $\gamma$ are arbitrary parameters.  As a consequence, $R_{11} = 0$, $R_{22} = 0$ and $R_{33} = 0$ reduce to

\begin{equation}
\label{R11 = R22= R33}
\left(\epsilon e^{\omega} + \xi^2 e^{\nu}\right)\left[\frac{f_{\xi \xi}}{f_{\xi}}+ \frac{f_{\xi}}{f}\left(\alpha + \beta + \gamma - 1\right)\right] - \frac{1}{2}\left(\nu_{\xi} - \omega_{\xi}\right)\left(\epsilon e^{\omega} - \xi^2 e^{\nu}\right) + 2\xi e^{\nu} = 0.
\end{equation}
On the other hand $R_{04}$ yields

\begin{equation}
\label{R04 in terms of f}
(\alpha + \beta + \gamma)\left(\frac{2 f_{\xi \xi}}{f_{\xi}} - \nu_{\xi} -  \omega_{\xi} + \frac{2}{\xi}\right) + 2(\alpha^2 + \beta^2 + \gamma^2 - \alpha - \beta - \gamma)\left(\frac{f_{\xi}}{f}\right) = 0,
\end{equation}
from which we get 
\begin{equation}
\label{integration of R04}
e^{(\nu + \omega)/2} = E \;\xi f^{(b/a)}f_{\xi},
\end{equation}
where $E$ is a constant of integration, and
\begin{equation}
\label{def. of a and b}
a \equiv (\alpha + \beta + \gamma), \;\;\;b \equiv (\alpha^2 + \beta^2 + \gamma^2 - \alpha - \beta - \gamma).
\end{equation}
Clearly, from (\ref{integration of R04}) it follows that there are different scenarios depending on whether $(\nu + \omega) \neq 0$ or $(\nu + \omega) = 0$. We now proceed to discuss them separately.

\subsubsection{General solution for $l = 2$}

For the general case where $\nu + \omega \neq 0$ in (\ref{integration of R04}), we substitute $e^{\nu/2} = E \xi f^{(b/a)}f_{\xi}e^{- \omega/2}$ into (\ref{R00}) and (\ref{R44}). Since  these are second order differential equations for $\nu$, we obtain two differential equations containing $f_{\xi \xi \xi}$, the third derivative of $f(\xi)$. Next, we isolate the $f_{\xi \xi \xi}$ obtained from $R_{00} = 0$ and $R_{44} = 0$, say $f_{\xi \xi \xi}(R_{00})$ and  $f_{\xi \xi \xi}(R_{44})$, respectively. Then, from the compatibility condition
\begin{equation}
\label{compatibility condition}
f_{\xi \xi \xi}(R_{00}) = f_{\xi \xi \xi}(R_{44}), 
\end{equation}
we find
\begin{equation}
\label{consequences of the compatibility condition}
e^{2 \omega} = - \epsilon E^2 \xi^4 f^{(2b/a)} f_{\xi}^2. 
\end{equation}
This expression demands  the extra dimension $\psi$ to be spacelike, i.e.,  $\epsilon = - 1$. In addition from (\ref{integration of R04}) we find
\begin{equation}
\label{Enu and Eom for the model under study}
e^{\nu} = E f^{(b/a)}f_{\xi}, \;\;\;\;\mbox{and} \;\;\;e^{\omega(\xi)} = \xi^2 e^{\nu(\xi)}.
\end{equation}
Now, it is easy to verify that the field equation (\ref{R11 = R22= R33}) is identically satisfied for $e^{\omega} = \xi^2 e^{\nu}$ and $\epsilon = - 1$. 

\medskip 

Collecting results, we have showed that the $5D$ line element 
\begin{equation}
\label{in summary}
dS^2 = E f^{(b/a)}f_{\xi}dt^2 - A f^{2 \alpha}dx^2 - B f^{2 \beta}dy^2 - C f^{2 \gamma}dz^2 - E \xi^2 f^{(b/a)}f_{\xi}d\psi^2,
\end{equation} 
is the general  solution of the field equations (\ref{R00})-(\ref{R04}), for {\it any} arbitrary function $f = f(\xi)$ with $\xi = t/\psi$. 

\subsubsection{Particular solution for $l = 2$}

For the case where $\nu + \omega =  0$, but $\nu = - \omega \neq 0$, integrating (\ref{integration of R04}) we obtain the solution as follows  
\begin{equation}  
\label{solution with nu = - omega}
dS^2 = \left(\frac{1}{\xi}\right)dt^2 - A h^{2 \alpha}dx^2 - B h^{2 \beta}dy^2 - C h^{2 \gamma}dz^2 - \xi d\psi^2,\;\;\;\mbox{with}\;\;\;h = [C_{1}\ln \xi + C_{2}]^{a/(a + b)},
\end{equation}
where $C_{1}$ and $C_{2}$ are constants of integration. It should be mentioned that this is the {\it unique} 
family of solutions with  $\nu = - \omega \neq 0$. 

In the case where $\nu = - \omega =  0$ the field equations $R_{AB} = 0$ require $a = \alpha + \beta + \gamma = 0$, i.e., $h = 1$. Thus, this case yields flat Minkowski space in $5D$ and $4D$. 

\subsection{Solutions for $l \neq 1$} 

In order to construct the solutions with $l \neq 1$, we now use the transformation (\ref{transforming the solutions from l = 2 to any l}). The second equation in (\ref{Enu and Eom for the model under study}) becomes 
\begin{equation}
\label{relation between Enu and Eomega for l neq 1}
e^{\omega(\zeta)} = \zeta^{2(l - 1)}e^{\nu(\zeta)}.
\end{equation}
On the other hand $f_{\xi} \rightarrow \zeta^{2 - l} f_{\zeta}$.
Thus, from (\ref{in summary}) we obtain the \underline{general solution for $l \neq 1$} as 
\begin{equation}
\label{solution for l neq 1, 2}
dS^2 = \bar{E} \zeta^{2 - l}f^{(b/a)}f_{\zeta}dt^2 - A f^{2 \alpha}dx^2 - B f^{2 \beta}dy^2 - C f^{2 \gamma}dz^2 - \bar{E} \zeta^l f^{(b/a)}f_{\zeta}d\psi^2. 
\end{equation}
In addition, from (\ref{solution with nu = - omega}) we get 
\begin{equation}  
\label{general solution with nu = - omega}
dS^2 = \left(\frac{1}{\zeta^{l - 1}}\right)dt^2 - A h^{2 \alpha}dx^2 - B h^{2 \beta}dy^2 - C h^{2 \gamma}dz^2 - \zeta^{l - 1} d\psi^2,\;\;\;\mbox{with}\;\;\;h \equiv  [C_{1}\ln \zeta + C_{2}]^{a/(a + b)},
\end{equation}
which in terms of $t$ and $\psi$ is identical to (\ref{solution with nu = - omega}). However, we put it here in explicit form in order to develop solutions with $l = 1$.

\subsection{Solutions for $l = 1$}

The field equations $R_{AB} = 0$ require $p = q$. Therefore the similarity variable now becomes 
\begin{equation}
\label{similarity variable for l= 1}
\zeta_{(l = 1)} \equiv \tilde{\zeta} =  \left(\frac{e^{t}}{e^{\psi}}\right)^q.
\end{equation}
In this case the solutions are readily obtained from (\ref{solution for l neq 1, 2}) and (\ref{general solution with nu = - omega}) just by setting $l = 1$ and replacing $\zeta$ by $\tilde{\zeta}$. Thus, from (\ref{relation between Enu and Eomega for l neq 1}) we get 
\begin{equation}
\label{relation between Enu and Eomega for l = 1}
e^{\omega(\tilde{\zeta})} = e^{\nu(\tilde{\zeta})}.
\end{equation}
Consequently, the \underline{general solution with $l = 1$} is given by 
\begin{equation}
\label{general solution for l = 1}
dS^2 = {E} \tilde{\zeta} f^{(b/a)}f_{\tilde{\zeta}}dt^2 - A f^{2 \alpha}dx^2 - B f^{2 \beta}dy^2 - C f^{2 \gamma}dz^2 - E \tilde{\zeta} f^{(b/a)}f_{\tilde{\zeta}}d\psi^2.
\end{equation}
Likewise,  from  (\ref{general solution with nu = - omega}) we get the particular solution
\begin{equation}  
\label{particular  solution with nu = - omega for l = 1}
dS^2 = dt^2 - A h^{2 \alpha}dx^2 - B h^{2 \beta}dy^2 - C h^{2 \gamma}dz^2 - d\psi^2,\;\;\;\mbox{with}\;\;\;h \equiv  [C_{1}\ln {\tilde{\zeta}} + C_{2}]^{a/(a + b)}.
\end{equation}

\subsection{Solutions for  $ a = \alpha + \beta + \gamma = 0$}

It should be noted that  (\ref{in summary}), (\ref{solution for l neq 1, 2}) and (\ref{general solution for l = 1}) require $a = (\alpha + \beta + \gamma) \neq 0$.  If $a = 0$, then from (\ref{R04 in terms of f}) it follows that either $b = 0$, or $f$ = constant. In both cases the $5D$ metric becomes
\begin{equation}
\label{a = 0}
dS^2 = e^{\nu(\zeta)}dt^2 - dx^2 - dy^2 - dz^2 + \epsilon e^{\omega(\zeta)}d\psi^2.
\end{equation}
At this point, it is crucial to emphasize that (\ref{relation between Enu and Eomega for l neq 1}) and (\ref{relation between Enu and Eomega for l = 1}) hold for any particular value of $\alpha, \beta $ and $\gamma$. Therefore, the line elements 
\begin{equation}
\label{a = 0, l neq 1}
dS^2 = e^{\nu(\zeta)}dt^2 - dx^2 - dy^2 - dz^2 - \zeta^{2(l - 1)}e^{\nu(\zeta)}d\psi^2,\;\;\;\;\zeta = \left(\frac{t}{\psi}\right)^{1/(l - 1)}, 
\end{equation}
and 
\begin{equation}
\label{a = 0, l = 1}
dS^2 = e^{\nu(\tilde{\zeta})}dt^2 - dx^2 - dy^2 - dz^2 - e^{\nu(\tilde{\zeta})}d\psi^2,\;\;\;\;\tilde{\zeta} = \left(\frac{e^{t}}{e^{\psi}}\right)^{q},
\end{equation}
constitute the most general solutions of the field equations $R_{AB} = 0$, for the metric (\ref{a = 0}). We also note that (\ref{solution with nu = - omega}), (\ref{general solution with nu = - omega}) and (\ref{particular  solution with nu = - omega for l = 1}) require $(a + b) = \alpha^2 + \beta ^ 2 + \gamma^2 \neq 0$. This is always satisfied unless $\alpha = \beta = \gamma = 0$, in which case the line element has the form (\ref{a = 0}).

\section{Some properties of the solutions}

The general solutions (\ref{solution for l neq 1, 2})  and (\ref{general solution for l = 1}) depend on one arbitrary function, of the corresponding similarity variable, and contain {\it three} arbitrary parameters $\alpha$, $\beta$ and $\gamma$. In addition, the field equations require the extra dimension to be spacelike $(\epsilon = - 1)$.
For $\alpha = \beta = \gamma$ we recover cosmological models with spatial spherical symmetry. Indeed,   
setting $f^{2 \alpha} = f^{2 \beta} = f^{2 \gamma} = e^{\lambda(\xi)}$; changing coordinates: $x = r \sin\theta \cos\phi$, $y = r \sin \theta\sin \phi$, $z = r \cos \theta$; re-naming the constants, without loss of generality  (\ref{solution for l neq 1, 2}) can be  written as 
\begin{equation}
\label{spherical solution with l neq 1}
dS^2 = \left(\frac{1}{C}\right)\zeta^{(2 - l)}\lambda_{\zeta}e^{\lambda(\zeta)/2}\;dt^2 - e^{\lambda(\zeta)}\left[dr^2 + r^2\left(d\theta^2 + \sin^2\theta d\phi^2\right)\right] - \left(\frac{1}{C}\right)\zeta^l\lambda_{\zeta}e^{\lambda(\zeta)/2}\;d\psi^2.  
\end{equation}
Likewise (\ref{general solution for l = 1}) becomes
\begin{equation}
\label{spherical solution with l = 1}
dS^2 = \left(\frac{1}{C}\right)\tilde{\zeta}\lambda_{\tilde{\zeta}}e^{\lambda(\tilde{\zeta})/2}\;dt^2 - e^{\lambda(\tilde{\zeta})}\left[dr^2 + r^2\left(d\theta^2 + \sin^2\theta d\phi^2\right)\right] - \left(\frac{1}{C}\right)\tilde{\zeta}\lambda_{\tilde{\zeta}}e^{\lambda(\tilde{\zeta})/2}\;d\psi^2.  
\end{equation}
These two metrics represent the most general self-similar, homogeneous and isotropic, Ricci-flat cosmologies in $5D$ \cite{UniverseTopSep}.

Although solutions (\ref{general solution for l = 1}) are formally obtained from (\ref{solution for l neq 1, 2}) just  by setting $l = 1$, they have different geometrical properties. In particular, solutions (\ref{solution for l neq 1, 2}) admit a homothetic Killing vector in $5D$ for {\it any} values of $\alpha$, $\beta$ and $\gamma$, namely,
\begin{equation}
\label{Lie derivative}
{\cal{L}}_{\zeta}{g_{A B}^{(l \neq 1)}} = 2 g_{A B}^{(l \neq 1)}, \;\;\;\mbox{with}\;\;\;\zeta^{C}_{(l \neq 1)} = (t, \; x, \; y, \; z, \; \psi),
\end{equation}
where $g_{AB}^{(l \neq 1)}$ is the metric (\ref{solution for l neq 1, 2}) and  ${\cal{L}}_{\zeta}$ denotes the Lie derivative along the $5D$ vector $\zeta^{C}_{(l \neq 1)}$. On the other hand, solutions (\ref{general solution for l = 1}) are self-similar but do {\it not} admit a homothetic Killing vector, except for $\alpha = \beta = \gamma$. In general, they admit an infinitesimal conformal transformation  parameterized by some function $H = H (t + \psi)$, viz.,
\begin{equation}
\label{conformal symmetry}
 {\cal{L}}_{\zeta}{g_{A B}^{(l = 1)}} = 2 \dot{H}g_{A B}^{(l = 1)}, \;\;\;\mbox{with}\;\;\;\zeta^{C}_{(l = 1)} = (H, \; x\dot{H}, \; y\dot{H}, \; z\dot{H}, \; H),\;\;\;\mbox{and}\;\;\; H = H(t + \psi).
\end{equation}

Another difference is  that (\ref{solution for l neq 1, 2}) is invariant under the transformation: $t \rightarrow i \psi$, $\psi \rightarrow i t$ (double Wick rotations) and  $\bar{E} \rightarrow - \hat{E}$. Indeed, it is easy to verify that 
\begin{equation}
\label{solution for l neq 1, 2 after Wick rotation}
dS^2 = \hat{E} \hat{\zeta}^lf^{(b/a)}f_{\hat{\zeta}}dt^2 - A f^{2 \alpha}dx^2 - B f^{2 \beta}dy^2 - C f^{2 \gamma}dz^2 - \hat{E} \hat{\zeta}^{(2 - l)} f^{(b/a)}f_{\hat{\zeta}}d\psi^2, \;\;\;\mbox{with}\;\;\;\hat{\zeta} \equiv \frac{1}{\zeta} 
\end{equation}
is also a solution of the field equations. The same occurs with (\ref{general solution for l = 1}) but the metric functions in the ``Wick rotated" line element are now complex functions.

However, the solutions share some important properties. For example,  one can show that there are only two cases where the components of the $5D$ Riemann tensor vanish: (i) $f =$ constant (or $\alpha = \beta = \gamma = 0$), and (ii) $\alpha = \beta = \gamma \neq 0$. 
In the first case the line element  and the  metric functions are given by (\ref{a = 0}). 
In the second case the spatial sections $t$ = constant and $\psi$ = constant are flat and possess spherical symmetry \cite{UniverseTopSep}. In any other circumstance, the manifold is curved in $5D$. As an illustration we present here the Riemann tensor in $5D$, calculated with the line element (\ref{in summary}),
\begin{eqnarray}
\label{Riemann tensor for l = 2}
R_{0114} &=& \xi R_{0101} = \frac{R_{1414}}{\xi} =  -  \frac{A \alpha f^{2(\alpha - 1)}\xi f_{\xi}^2 \left[\beta^2 + \gamma^2 - \alpha(\beta + \gamma)\right]}{a \psi^2},  \nonumber \\
R_{0224} &=& \xi R_{0202} = \frac{R_{2424}}{\xi} =  -  \frac{B \beta f^{2(\beta - 1)}\xi f_{\xi}^2 \left[\alpha^2 + \gamma^2 - \beta(\alpha + \gamma)\right]}{a \psi^2}, \nonumber \\
R_{0334} &=& \xi R_{0303} = \frac{R_{3434}}{\xi} =  - \frac{C \gamma f^{2(\gamma - 1)}\xi f_{\xi}^2 \left[\alpha^2 + \beta^2 - \gamma(\alpha + \beta)\right]}{a \psi^2}.
\end{eqnarray}
Finally, it is essential to mention that  the solutions discussed in the preceding section are invariant under the change $(x, \; y, \; z) \leftrightarrow \psi$, which  is a consequence of the fact that the extra coordinate $\psi $ is spacelike.  For example, if we change $\psi \leftrightarrow z$ and denote
\begin{equation}
\label{def. of eta}
F \equiv F(\eta), \;\;\;\;\mbox{with}\;\;\;\eta = \frac{t}{z},
\end{equation} 
then the  line element
\begin{equation}
\label{interchanging x and psi}
dS^2 = E F^{(b/a)}F_{\eta}\left(dt^2 - \eta^2 dz^2\right) - A F^{2 \alpha}dx^2 - B F^{2 \beta}dy^2 - CF^{2 \gamma}d\psi^2, 
\end{equation}
which we have constructed from  (\ref{in summary}), is also a solution of the field equations $R_{AB} = 0$. Certainly, the same is true for all solutions discussed in section $3$. It is important to note that the metrics obtained under the change $(x, \; y, \; z) \leftrightarrow \psi$ (i) are independent on the ``extra" dimension $\psi$,  and (ii) the spatial sections defined by $t = $ constant, $\psi =$ constant are non-flat. In the next section we will see that this symmetry is crucial for the interpretation of the solutions in $4D$.

\section{Interpretation in $4D$}

We have already mentioned that  there are different ways of producing, or embedding, a $4D$ spacetime in a given higher-dimensional manifold. However, the most popular approach is based on three different assumptions. First, that we can use the coordinate frame \cite{Coord. frame}. Second, that our $4D$ spacetime can be recovered by going onto a hypersurface $\Sigma_{\psi}: \psi =  \psi_{0} = $ constant, which is orthogonal to the $5D$ unit vector
\begin{equation}
\label{unit vector n}
{\hat{n}}^{A} = \frac{\delta^{A}_{4}}{\sqrt{\epsilon g_{44}}}, \;\;\;n_{A}n^{A} = \epsilon,
\end{equation}
along the extra dimension. Third, that the physical metric of the spacetime can be identified with  the one induced on $\Sigma_{\psi}$.  

For a line element of the form 
\begin{equation}
\label{General line element in 5D without off-diagonal terms}
dS^2 = g_{\mu\nu}(x^{\rho}, \; \psi)dx^{\mu}dx^{\nu} + \epsilon \Phi^2(x^{\rho}, \; \psi)d\psi^2,
\end{equation}
 the induced metric on hypersurfaces $\Sigma_{\psi}$ is just $g_{\mu\nu}$, i.e., the $4D$ part of the metric in $5D$. The crucial moment is that, although the energy-momentum tensor (EMT) in $5D$ is zero, to an observer  confined to making physical measurements in our ordinary spacetime, and not  aware of the extra dimension,  the spacetime is not empty but contains (effective) matter whose EMT,  ${^{(4)}T}_{\alpha\beta}$, is determined by the Einstein equations in $4D$, namely
\begin{eqnarray}
\label{4D Einstein with T and K}
{^{(4)}G}_{\alpha\beta} = 8 \pi \;{^{(4)}T}_{\alpha\beta} = 
- \epsilon\left(K_{\alpha\lambda}K^{\lambda}_{\beta} - K_{\lambda}^{\lambda}K_{\alpha\beta}\right) + \frac{\epsilon}{2} g_{\alpha\beta}\left(K_{\lambda\rho}K^{\lambda\rho} - (K^{\lambda}_{\lambda})^2 \right) - \epsilon E_{\alpha\beta}, 
\end{eqnarray}
where $K_{\mu\nu}$ is the extrinsic curvature 
\begin{equation}
\label{extrinsic curvature}
K_{\alpha\beta} = \frac{1}{2}{\cal{L}}_{\hat{n}}g_{\alpha\beta} = \frac{1}{2\Phi}\frac{\partial{g_{\alpha\beta}}}{\partial \psi};  
\end{equation}
$E_{\mu\nu}$ is the projection of the bulk Weyl tensor ${^{(5)}C}_{ABCD}$ orthogonal to ${\hat{n}}^A$, i.e., ``parallel" to spacetime, viz.,
\begin{equation}
\label{Weyl Tensor}
E_{\alpha\beta} = {^{(5)}C}_{\alpha A \beta B}{\hat{n}}^A{\hat{n}}^B 
= - \frac{1}{\Phi}\frac{\partial K_{\alpha\beta}}{\partial \psi} + K_{\alpha\rho}K^{\rho}_{\beta} - \epsilon \frac{\Phi_{\alpha;\beta}}{\Phi},
\end{equation}
and $\Phi_{\alpha} \equiv \partial \Phi/\partial x^{\alpha}$. It is important to mention that the effective matter content of the spacetime is the same whether we interpret it in space-time-matter theory \cite{Wesson book}-\cite{Wesson and JPdeL}, or in a ${\bf Z}_2$ symmetric brane universe \cite{Shiromizu}. Indeed, these two theories are mathematically equivalent, although they have different motivation and physical interpretation \cite{STM-Brane}.

\subsection{Homogeneous cosmological models}

Since $\zeta = \xi^{1/(l - 1)}$, it follows that the models with $l \neq 1$, discussed in Section $3.1$,  are physically indistinguishable from those with $l = 2$. The only reason for  considering them here is a mathematical one: they allow us to generate non-homothetic $5D$  models ($l = 1$) from homothetic ones ($l \neq 1$). In what remains we focus our attention to metrics (\ref{in summary}) and (\ref{interchanging x and psi}), which admit a homothetic Killing vector in $5D$, namely (\ref{Lie derivative}), and therefore are relevant to the similarity hypothesis mentioned in the Introduction\footnote{We shall come back to discuss the $4$-dimensional interpretation of the  non-homothetic $5D$ models $(l = 1)$ elsewhere.}. 

However, by virtue of the symmetry of the solutions in $5D$, which in the case under consideration is expressed by (\ref{solution for l neq 1, 2 after Wick rotation}) and  (\ref{interchanging x and psi}),   there are a number of distinct  possible scenarios in $4D$. An exhaustive treatment  of all these scenarios is  beyond the purview of this work.  Rather, in this section  we restrict ourselves to presenting some representative models.

In the approach under consideration, the spacetime metric generated by  (\ref{in summary}) is given by

\begin{equation}
\label{anisotropic cosmological metrics}
ds^2 \equiv dS^2_{|\Sigma_{\psi}} =  g_{\mu\nu}dx^{\mu}dx^{\nu} = E f^{(b/a)}f_{\xi}dt^2 - A f^{2 \alpha}dx^2 - B f^{2 \beta}dy^2 - C f^{2 \gamma}dz^2.
\end{equation}
Let us emphasize that, on each hypersurface $\Sigma_{\psi}$, this  metric is a function of time only. Therefore, (\ref{anisotropic cosmological metrics}) yields homogeneous, anisotropic cosmological models of   Bianchi type I.

 A general feature of metrics (\ref{anisotropic cosmological metrics}) is that they are partially homothetic, i.e.,
\begin{equation}
\label{spatial}
 {\cal{L}}_{\zeta}g_{00} =  0, \;\;\;\;{\cal{L}}_{\zeta}g_{i j} = 2 g_{i j}, \;\;\;\mbox{with}\;\;\;\zeta^{\mu} = (0, \; x, \; y, \;z),
\end{equation}
for any arbitrary function $f$. 

\subsubsection{Models with homothetic symmetry in $4D$}

The condition of homothetic symmetry in $4D$ singles out  a unique family of metrics, which are generated by the choice $f = \xi^{a/(a + b)}$ (See the Appendix). The homothetic $4D$ vector as well as the embedding solution in $5D$ are given by (\ref{homothetic 4D vector for the homogeneous model}) and (\ref{homothetic 5D and 4D metric}), respectively.
The $4D$ line element with homothetic symmetry is given by (\ref{homothetic solution in 4D metric}),
\begin{equation}
\label{homothetic solution in 4D metric, main text}
ds^2 = dt^2 - \bar{A}\;t^{2\alpha k}dx^2 - \bar{B}\;t^{2\beta k}dy^2 - \bar{C}\;t^{2\gamma k}dz^2,\;\;\;k \equiv \frac{a}{a + b} = \frac{\alpha + \beta + \gamma}{\alpha^2 + \beta^2 + \gamma^2}.
\end{equation}
We note that the volume of the spatial slices goes like $t^{ak}$. Since $ak > 0$, it follows that the volume of space is increasing from an original big-bang at $t = 0$. This is regardless of whether some particular direction is contracting.   
The effective EMT is\footnote{To simplify the notation, in what follows we suppress the index $^{(4)}$ in ${^{(4)}T}_{\alpha\beta}$.} 
\begin{equation}
\label{ETM for homothetic homogeneous metric}
8 \pi T_{0}^{0} = \frac{k^2 (\alpha \beta + \alpha \gamma + \beta \gamma)}{t^2}, \;\;\;\;
T_{1}^{1} = n_{x}T_{0}^{0}, \;\;\;\;T_{2}^{2} = n_{y}T_{0}^{0}, \;\;\;\;
T_{3}^{3} = n_{z}T_{0}^{0},
\end{equation}
where $n_{x}$, $n_{y}$ and $n_{z}$ are constants given by
\begin{equation}
\label{the constants nx, ny and nz}
n_{x} = \frac{(- \alpha + \beta + \gamma)}{\alpha + \beta + \gamma}, \;\;\;\;n_{y} = \frac{(- \beta + \alpha + \gamma)}{\alpha + \beta + \gamma}, \;\;\;\;n_{z} = \frac{(- \gamma + \alpha + \beta)}{\alpha + \beta + \gamma}.
\end{equation}
We note that 
\begin{equation}
\label{relationship between the n's}
n_{x} + n_{y} + n_{z} = 1,
\end{equation}
for any value of $\alpha$, $\beta$ and $\gamma$. Let us notice some particular cases:  (I) If one of the coefficients vanishes, say $\alpha = 0$, then $n_{x} = 1$ and $n_{y} = - n_{z}$; (II) If two coefficients are equal, say $\alpha = \beta$, then $n_{x} = n_{y}$. In particular, if $\alpha = \beta = 0$, then $n_{x} = n_{y} = 1$ and $n_{z} = -1$; (III) If $\alpha = \beta = \gamma \neq 0$, then $n_{x} = n_{y} = n_{z} = 1/3$.  The homothetic model exhibits a number of interesting properties, however a detailed study of the matter sources that satisfy 
(\ref{ETM for homothetic homogeneous metric})-(\ref{relationship between the n's}) is  outside the scope of the present work. 

\subsubsection{Perfect fluid with stiff equation of state}
The function  $f(\xi)$ in (\ref{anisotropic cosmological metrics}) can be determined by imposing some ``physical" conditions on the effective picture in  $4D$. If we assume that the effective matter behaves like a perfect fluid, then we obtain 
\begin{equation}
\label{f(xi) for the solution with stiff equation}
f(\xi) = \left(C_{1}\xi + C_{2}\right)^{m}, \;\;\;\;\mbox{where}\;\;\;\;m \equiv \frac{\alpha + \beta + \gamma}{\alpha^2 + \beta^2 + \gamma^2 + 4(\alpha \beta + \alpha \gamma + \beta \gamma)},
\end{equation}
and $C_{1}, C_{2}$ are some constants. The effective density $\rho = T_{0}^{0}$ and pressure $p  = 
- T_{1}^{1} = - T_{2}^{2} = - T_{3}^{3}$ in $4D$ satisfy the stiff equation of state, namely,
\begin{equation} 
\label{stiff equation}
\rho = p = \frac{\left[\gamma(\alpha + \beta) + \alpha \beta\right]m \; C_{1}}{8 \pi E \psi^2 f^{2(\alpha + \beta + \gamma)}}.
\end{equation}

\subsubsection{Homothetic $5D$ embedding for the $4D$ Kasner spacetime}

For completeness, we should mention that on each spacetime section $\Sigma_{\psi}$ we have
\begin{equation}
\label{T00 for homogeneous models}
8 \pi T_{00} = \frac{\gamma (\alpha + \beta) + \alpha \beta}{\psi^2}\left(\frac{f_{\xi}}{f}\right)^2.
\end{equation}
Therefore, vacuum solutions in $4D$ require $\gamma = - \alpha \beta/(\alpha + \beta)$. Then, from the vanishing of $T_{ij}$ we obtain
  
\begin{equation}
\label{homogeneous vacuum solutions } 
f(\xi) = \left(C_{1}\xi + C_{2}\right)^{\frac{\alpha + \beta}{\alpha^2 +  \alpha \beta + \beta^2}},  
\end{equation}
where $C_{1}$ and $C_{2}$ are constants of integration.
Now, setting  $C_{1} = [(\alpha^2 + \alpha \beta + \beta^2)/E(\alpha + \beta)]$, we obtain  $g_{00} = 1$. Thus, the metric in $5D$ becomes

\begin{equation}
\label{5D self-similar embedding for Kasner}
dS^2 = dt^2 - A(C_{1}\xi + C_{2})^{2 p_{1}}dx^2 - B(C_{1}\xi + C_{2})^{2 p_{2}}dy^2- C(C_{1}\xi + C_{2})^{2 p_{3}}dz^2 - \xi^{2 p_{4}}d\psi^2,
\end{equation}
where
\begin{equation}
\frac{p_{1}}{\alpha(\alpha + \beta)} = \frac{p_{2}}{\beta(\alpha + \beta)}= - \frac{ p_{3}}{\alpha \beta} = \frac{1}{\alpha^2 + \alpha \beta + \beta^2}; \;\;\;p_{4} = 1.
\end{equation}
We note that
\begin{equation}
p_{1} + p_{2} + p_{3} = 1, \;\;\;\; p_{1}^2 + p_{2}^2 + p_{3}^2  = 1. 
\end{equation}
Therefore, (\ref{5D self-similar embedding for Kasner}) can be interpreted as a homothetic $5D$ embedding for the $4D$ Kasner spacetime. It is important  to point out   two things. Firstly, that although (\ref{5D self-similar embedding for Kasner}) represents an empty space in $5D$, it is {\it not} equivalent to the usual  $5$-dimensional Kasner space.\footnote{We recall the reader that a Kasner space in $5D$  is given by $dS^2 = dt^2 - At^{2 p_{1}}dx^2 - B t^{2 p_{2}}dy^2 - C t^{2 p_{3}}dz^2 + \epsilon E t^{{2 p_{4}}} d\psi^2$ with $p_{1} + p_{2} + p_{3} + p_{4}= 1, \;\;\;\; p_{1}^2 + p_{2}^2 + p_{3}^2  + p_{4}^2= 1$.} Secondly, it illustrates that the effective EMT  in $4D$, which is given by (\ref{4D Einstein with T and K}) can vanish even though the induced metric in $4D$ explicitly depends on the extra coordinate. This is totally different from the isotropic cosmological models, for which the effective EMT is always nonzero when the $5D$ metric has a explicit dependence on the extra coordinate. In fact, this is a general attribute of models with spatial spherical symmetry \cite{PdeLWesson}.

\subsection{Inhomogeneous models: nonstatic vacuum solutions}

Let us now consider the $5D$ line element (\ref{interchanging x and psi}). In the approach under consideration, the spacetime metric is given by
\begin{equation}
\label{interchanging x and psi, spacetime metric}
ds^2 \equiv dS^2_{|\Sigma_{\psi}} = g_{\mu\nu} dx^{\mu}dx^{\nu} = E F^{(b/a)}F_{\eta}\left(dt^2 - \eta^2 dz^2\right) - A F^{2 \alpha}dx^2 - B F^{2 \beta}dy^2, 
\end{equation}
where $F$ is a function of the self-similar variable $\eta = t/z$. These models present two main characteristics. First, they show homothetic symmetry along the $4D$ vector
\begin{equation}
\label{spacetime projection of the homothetic vector in 5D}
\xi^{\mu}_{p} = (t, \; x, \; y, \; z),
\end{equation}
where the subindex $p$ indicates that $\xi^{\mu}_{p}$ is the spacetime projection of the $5D$ homothetic vector $\xi^{A}$ defined in (\ref{Lie derivative}). Second, since the spacetime part of the generating $5D$ metric (\ref{interchanging x and psi}) is independent of $\psi$, it follows  that $K_{\mu\nu} = 0$ and consequently the effective EMT is 
\begin{equation}
\label{EMT for the inhomogeneous models}
8 \pi T_{\mu\nu} = E_{\mu\nu} = \frac{\Phi_{\alpha; \beta}}{\Phi},
\end{equation} 
we recall that the general solution requires $\epsilon = - 1$. Therefore,
\begin{equation}
T \equiv T_{0}^{0} + T_{1}^{1} + T_{2}^{2} + T_{3}^{3} = 0,
\end{equation}
which is a consequence of the fact that $E_{\mu\nu}$ is traceless.

\subsubsection{Solutions in conventional $4D$ general relativity}

The nonvanishing components of the EMT are
\begin{equation}
\label{EMT for the RS vacuum solutions}
T_{0}^{0} = - T_{3}^{3} \;= - \left(\frac{z}{t}\right)T_{3}^{0} =  \frac{\gamma \left[\gamma \left(\alpha + \beta\right) - \alpha^2 - \beta^2\right]}{8 \pi(\alpha + \beta + \gamma)\;E}\left[\frac{F_{\eta}}{z^2\; F^{(2 a + b)/a}}\right]. 
\end{equation}
They can be interpreted as ``pancake-like" distributions of matter, with energy flow  along the direction of symmetry $z$. A detailed study of the matter distribution (\ref{EMT for the RS vacuum solutions}) is beyond the scope of the present work.

For another interpretation, we note that $T_{\mu\nu} = 0$ when 
\begin{equation}
\label{choice of gamma for vacuum solutions}
\gamma = \frac{\alpha^2 + \beta^2}{\alpha + \beta},\;\;\;\alpha \neq - \beta.
\end{equation}
With this choice, the line element (\ref{interchanging x and psi, spacetime metric}) yields a two-parameter family of non-static inhomogeneous vacuum solutions in conventional $4D$ general relativity. For these solutions, the non-vanishing components of the Riemann tensor in $5D$ and $4D$ are all proportional       to $(\alpha - \beta)F_{\eta}$.  Consequently, for $\alpha \neq \beta$ the vacuum solutions given by (\ref{interchanging x and psi, spacetime metric}) and (\ref{choice of gamma for vacuum solutions}) are {\it not} equivalent to a $4D$ Minkowski spacetime and {\it cannot} be embedded in a Riemann-flat manifold in $5D$.

\subsubsection{Vacuum solutions in the Randall-Sundrum (RS2) braneworld scenario}

Without going into deep technical details, in the  Randall $\&$ Sundrum  braneworld scenario \cite{Randall2} our universe is identified with a {\it fixed } singular hypersurface $\Sigma_{\psi_{b}}$ (the {\it brane}) embedded in a   $5$-dimensional bulk  with ${\bf Z}_{2}$ symmetry  with respect to $\Sigma_{\psi_{b}}$\cite{Shiromizu}. Due to the presence of matter on the brane, which is  described by an EMT that we denote as $\tau_{\mu\nu}$,  the extrinsic curvature $K_{\mu\nu}$ is discontinuous across $\Sigma_{\psi_{b}}$. Israel's boundary conditions \cite{Israel} relate the jump of $K_{\mu\nu}$ to $\tau_{\mu\nu}$, namely, $({K_{\mu \nu}}_{|\Sigma^{+}_{\psi_{b}}} - {K_{\mu \nu}}_{|\Sigma^{-}_{\psi_{b}}}) = k (\tau_{\mu\nu} - \frac{1}{3}\tau g_{\mu\nu})$, where $k$ is a constant with the appropriate units. Next, ${\bf Z}_{2}$ symmetry implies 
${K_{\mu \nu}}_{|\Sigma^{+}_{\psi_{b}}} = - {K_{\mu \nu}}_{|\Sigma^{-}_{\psi_{b}}} \equiv {K_{\mu \nu}}$ . Therefore,  when the spacetime (the brane) is empty ($\tau_{\mu\nu} = 0$) it follows that $K_{\mu\nu} = 0$. Thus, in braneworld theory the vacuum field equations are obtained from (\ref{4D Einstein with T and K}) as
\begin{equation}
\label{vacuum field equations in braneworld}
{^{(4)}G}_{\alpha\beta} = 8 \pi \;{^{(4)}T}_{\alpha\beta} =  - \epsilon E_{\alpha\beta},
\end{equation}
which means that, in braneworld theory, the effective {\it geometrical} matter is $4D$ is traceless. Therefore, any solution of general relativity with $T = 0$ is a solution of the vacuum equations in braneworld.

The conclusion is that (\ref{interchanging x and psi, spacetime metric})  can be interpreted as vacuum solutions in the Randall-Sundrum braneworld scenario.

\section{Summary}

Self-similar (homothetic) symmetry seems to be very important in nature\footnote{We would like to share with our  readers a beautiful quotation about self-similar (homothetic) symmetry, ascribed to Manfred Schroeder, that we learned from a paper by Adrian Popesku \cite{Popesku}: ``The unifying concept underlying fractals,
chaos and power laws is self-similarity.
Self-similarity, or invariance against changes
in scale or size, is an attribute of many
laws of nature and innumerable phenomena
in the world around us. Self-similarity
is, in fact, one of the decisive symmetries
that shapes our universe and our efforts
to comprehend it."}. 
In cosmological applications, within the context of conventional general relativity in $4D$, there is
a strong evidence that many homogeneous and inhomogeneous cosmological models  can  be approximated by self-similar models in the asymptotic
regimes,  i.e.,  near the initial cosmological singularity and at late times \cite{Coley}-\cite{Apostolopoulos 3}. In order to avoid misunderstanding, it is important to emphasize that in the literature the concept of self-similarity is frequently equated with homothetic symmetry. In this paper we have followed the traditional terminology used in \cite{Sedov}-\cite{Barenblat}, where self-similarity means that the field equations are functions only of a single variable,  which in turn allows to reduce them to a system of ordinary differential equations. 

Observations indicate that on large scales ($\gg 100$ Mpc) the  universe is homogeneous and isotropic and well described by spatially-flat FRW cosmologies, which are self-similar in $4D$ and $5D$. 
However, there is no reason to expect such features   at the early stages of the evolution of the universe. Rather, it is generally accepted that anisotropy could have played a significant role in the early universe and that it has been fading away in the course of cosmic evolution. Therefore, the study of cosmological models that are anisotropic and self-similar appears to be of primary interest.

In this work we have investigated such cosmological models within the context  of theories of Kaluza-Klein type, with a large extra dimension. The ladder to go   between the physics in $4D$ and $5D$ is provided by Campbell-Maagard's embedding theorems,  which guarantee that any solution of the $4D$ Einstein equations of general relativity may be embedded in a solution of the $5D$ vacuum Einstein equations.

For the case where the $5D$ metric is diagonal and is a function only   of time and the ``extra" coordinate $\psi$, we have shown that there are  three possible forms for the  similarity variable: $\xi = t/\psi$; $\zeta = (e^{t}/e^{\psi})^{q}$, and $\zeta = \omega_{0} t + k_{0}\psi$. They correspond to three different physical situations, which are described by  homothetic, conformal and plane wave-like solutions in $5D$, respectively.  Thus, in the traditional terminology the concept of self-similarity includes conformal and wave-like solutions.

In section $3$, we have constructed the most general homothetic and conformal solutions in $5D$ to the field equations (\ref{R00})-(\ref{R04}). They are given in terms of one arbitrary function, of the appropriate similarity variable, and three arbitrary parameters, viz., $\alpha, \beta$, and $\gamma$. However, the situation is completely different for the case where the $5D$ metric is wave-like. This case is more involved and requires a separate discussion.  

Following Campbell-Maagard's embedding theorems, the connection to $4D$ is deduced after choosing an embedding\footnote{As it was mentioned in the Introduction, the physically interesting case of finding a $5D$ embedding for a given $4D$ spacetime, with a specified physical EMT, is a very difficult  task because the effective $4D$ equations for gravity  contain a source term, that is the spacetime projection of the $5D$ Weyl tensor, which is unknown without specifying the properties of the metric in $5D$ \cite{Maartens}.}. We have found that the solutions that are homothetic in $5D$ are relevant to the similarity hypothesis because all the  models constructed in $4D$, by means of dimensional reduction of the metric  in $5D$,  exhibit  some type of self-similarity. Specifically, we have seen that  
those that are inhomogeneous and anisotropic inherit the homothetic symmetry from the $5D$ embedding. The rest of them are either partially homothetic or homothetic along some $4D$ vector. 

Our discussion in section $5$  illustrates two important things. Firstly, that  when  the extra dimension is spacelike the concepts  of $4D$ spatial homogeneity  and $4D$ spatial flatness become totally dependent on the embedding, i.e., on how the coordinates for our $3$-dimensional space are chosen. Secondly, that the interpretation in $4D$ crucially depends on the theory under consideration. In consequence,  much work is still needed in order to understand $4D$ physical models as Lorentzian hypersurfaces in pseudo-Riemannian $5D$ spaces.

In summary: our work (I) generalizes the $5D$ embeddings used for the FLRW models; (II) shows that 
anisotropic cosmologies are, in general,  curved in $5D$,    in contrast with FLRW models which can always be embedded in a $5D$ Riemann-flat (Minkowski) manifold; (III) reveals that anisotropic  cosmologies  can be curved and devoid of matter, both   in $5D$ and $4D$, even when the metric in $5D$ explicitly depends on the extra coordinate, which is quite different from isotropic cosmological models where the effective EMT is always nonzero when the $5D$ metric has a explicit dependence on the extra coordinate. In fact, this is a general attribute of models with spatial spherical symmetry \cite{PdeLWesson}.

To finish this paper,  we should  mention that from a mathematical point of view the shape of our solutions can be simplified if we introduce ``null-like" coordinates (We thank Philippe Spindel for pointing this out \cite{Spindel}). As an illustration, let us consider (\ref{in summary}). The $(t-\psi)$ part of that metric can be factorized and written as  
$E f^{b/a}f_{\xi}(dt + \xi d\psi)(dt - \xi d\psi) \equiv 2 du dv$. Then, setting $u = f^{(a + b)/a}$ the line element (\ref{in summary}) becomes

\begin{equation}
\label{in summary in null coordinates}
dS^2 = 2 du dv - A u^{p_{1}}dx^2 - B u^{p_{2}}dy^2 - C u^{p_{3}}dz^2,
\end{equation} 
where  $p_{1} = 2\alpha a/(a + b)$, $p_{2} = 2\beta  a/(a + b)$ and $p_{3} = 2\gamma  a/(a + b)$. We note that $\sum_{i = 1}^{3}{(p_{i} - 1)^2 = 3}$. Although in this coordinates the solution looks mathematically simpler than in the original form, its physical interpretation in $4D$ is elusive since it not clear how to define the hypersurfaces $\Sigma_{\psi}$ orthogonal to the $5D$ unit vector ${\hat{n}}^{A}$ along the extra dimension. 
Therefore, in these coordinates neither the dimensional reduction discussed in Section $5$ nor Campbell's theorem can be used for the $4D$ interpretation of the solutions.

This investigation can be extended, or generalized, in different ways.  They follow from the fact that we have not fully examined the possible $4D$ interpretations of the $5D$-homothetic solutions (\ref{solution for l neq 1, 2}). Neither the conformal solutions (\ref{general solution for l = 1}), nor the plane wave-like solutions induced by the similarity variable (\ref{xi in the particular case}),  have been studied.  These are important topics that should be addressed.

\renewcommand{\theequation}{A-\arabic{equation}}
  \setcounter{equation}{0}  
  \section*{Appendix: Homothetic symmetry on $\Sigma_{\psi}$}  

Our aim here is to show that the requirement of homothetic symmetry on $\Sigma_{\psi}$ singles out one specific metric in $4D$. With this aim, let us take the Lie derivative of (\ref{anisotropic cosmological metrics}) along the $4D$ vector 
\begin{equation}
\label{homothetic vector in 4D}
\zeta^{\mu}_{h} = \left(C_{0}\; t, \;  C_{1}\;x, \; C_{2}\; y, \; C_{3}\; z\right),
\end{equation}
where $h$ stands for ``homothetic" and $C_{0}, C_{2}, C_{3}, C_{3}$ are some constants. We obtain, 
\begin{eqnarray}
{\cal{L}}_{\zeta_{h}}g_{00} &=& 2 g_{00}\;C_{0}\left[1 + \frac{\xi}{2}\left(\frac{b f_{\xi}}{a f} + \frac{f_{\xi \xi}}{f_{\xi}}\right)\right], \nonumber \\
{\cal{L}}_{\zeta_{h}}g_{11} &=& 2 g_{11}\left[ C_{1} + \alpha C_{0} \xi\left(\frac{f_{\xi}}{f}\right) \right], \nonumber \\
{\cal{L}}_{\zeta_{h}}g_{22} &=& 2 g_{22}\left[ C_{2} + \beta C_{0} \xi \left(\frac{f_{\xi}}{f}\right) \right], \nonumber \\
{\cal{L}}_{\zeta_{h}}g_{33} &=& 2 g_{33}\left[ C_{3} + \gamma C_{0} \xi \left(\frac{f_{\xi}}{f}\right) \right].
\end{eqnarray}
The condition  ${\cal{L}}_{\zeta_{h}}g_{\mu \nu} = 2 g_{\mu \nu}$ requires
\begin{equation}
\frac{1 - C_{1}}{\alpha C_{0}} = \frac{1 - C_{2}}{\beta C_{0}} = \frac{1 - C_{3}}{\gamma C_{0}} \equiv k,\;\;\;\;\;f(\xi) \sim \xi^{k}\;\;\;\;\;\mbox{and}\;\;\; C_{0} = \frac{2 a}{a + k(a +b)},
\end{equation}
where $k$ is some constant. Without loss of generality we can set $C_{0} = 1$, which implies
\begin{equation}
\label{homothetic 4D vector for the homogeneous model}
k = \frac{a}{a + b}, \;\;\;\mbox{and}\;\;\;\zeta^{\mu}_{h} = \left[t,\;  (1 - k \alpha)x, \;(1 - k \beta)y, \; (1 - k \gamma)z\right].
\end{equation}
Thus, the $5D$ metric 
\begin{equation}
\label{homothetic 5D and 4D metric}
dS^2 = dt^2 - \xi^{2\alpha k}dx^2 - \xi^{2\beta k}dy^2 - \xi^{2\gamma k}dz^2 - \xi^2 d\psi^2,
\end{equation}
generates, on every $\Sigma_{\psi}$, the $4D$ line element 
 \begin{equation}
\label{homothetic solution in 4D metric}
ds^2 \equiv  dS^2_{|\Sigma_{\psi}} = dt^2 - \bar{A}\;t^{2\alpha k}dx^2 - \bar{B}\;t^{2\beta k}dy^2 - \bar{C}\;t^{2\gamma k}dz^2,
\end{equation}
 which shows homothetic symmetry along $\zeta^{\mu}_{h}$. Here we have introduced  the constants $\bar{A}$, $\bar{B}$ and $\bar{C}$ for dimensional consistency.  It should be noted that $\zeta^{\mu}_{h}$ is {\it not} the spacetime projection of the $5D$ vector $\zeta^{C}_{(l \neq 1)} = (t, \; x, \; y, \; z, \; \psi)$ defined in (\ref{Lie derivative}).

\end{document}